\documentclass[10pt,letterpaper]{article}
\usepackage[top=0.85in,left=1in,footskip=0.75in,marginparwidth=1in]{geometry}

\usepackage[utf8]{inputenc}

\usepackage{cite}

\usepackage{nameref}

\usepackage{microtype}
\DisableLigatures[f]{encoding = *, family = * }

\usepackage{changepage}
\usepackage{subfigure}

\usepackage[aboveskip=1pt,labelfont=bf,labelsep=period,singlelinecheck=off]{caption}

\makeatletter
\renewcommand{\@biblabel}[1]{\quad#1.}
\makeatother

\usepackage{lastpage,fancyhdr,graphicx}
\usepackage{epstopdf}
\fancyheadoffset[L]{2.25in}
\fancyfootoffset[L]{2.25in}
\usepackage{color}

\definecolor{Gray}{gray}{.25}

\usepackage{graphicx}

\usepackage{sidecap}

\usepackage{wrapfig}
\usepackage[pscoord]{eso-pic}
\usepackage[fulladjust]{marginnote}
\reversemarginpar

\begin{document}
\vspace*{0.35in}

\begin{flushleft}
{\Large
\textbf\newline{NetSim -- The framework for complex network generator}
}
\newline
\\
Akanda Wahid -Ul- Ashraf\textsuperscript{1,*},
Marcin Budka\textsuperscript{2},
Katarzyna Musial\textsuperscript{1},
\\
\bigskip
\bf{1,2} Department of Computing and Informatics, Bournemouth University Fern Barrow, Poole BH12 5BB, UK
\\
\bf{3} Advanced Analytics Institute, School of Software, Faculty of Engineering and IT, University of Technology Sydney, Australia
\\
\bigskip
* aashraf@bournemouth.ac.uk

\end{flushleft}

\section*{Abstract}
Networks are everywhere and their many types, including social networks, the Internet, food webs etc., have been studied for the last few decades. However, in real-world networks, it's hard to find examples that can be easily comparable, i.e. have the same density or even number of nodes and edges. 
We propose a flexible and extensible NetSim framework to understand how properties in different types of networks change with varying number of edges and vertices. Our approach enables to simulate three classical network models (random, small-world and scale-free) with easily adjustable model parameters and network size. To be able to compare different networks, for a single experimental setup we kept the number of edges and vertices fixed across the models. To understand how they change depending on the number of nodes and edges we ran over 30,000 simulations and analysed different network characteristics that cannot be derived analytically. Two of the main findings from the analysis are that the average shortest path does not change with the density of the scale-free network but changes for small-world and random networks; the apparent difference in mean betweenness centrality of the scale-free network compared with random and small-world networks.


\section{Introduction}
\label{Introduction}
Network properties such as closeness centrality, betweenness centrality, degree distribution, clustering coefficient, and average geodesic path vary depending on the network's type and size. These properties contain essential information about the network's structure and dynamics. For instance, a high clustering coefficient with low average geodesic path might indicate a social network, i.e. a specific type of network that is formed through social interactions~\cite{watts1998collective}. Degree distribution could indicate a particular growth mechanism unique to a specific kind of network -- power law in the degree distribution indicates that the growth mechanism is likely to be based on preferential attachment and the network has been formed based on the `rich get richer' rule~\cite{barabasi1999emergence}. To understand how different properties of a network behave with regards to the type of the network, and how properties change with network's size, we have designed and performed a number of simulations. In these simulations we have compared properties of three network models: 1)~Barab\'{a}si-Albert model for scale-free network 2)~Watts-Strogatz model for the small-world network, and 3)~Erd\H{o}s-R\'{e}nyi model for the random network.
To compare properties across three different network models of varying size and density, all these models for a particular size need the same number of vertices and edges. Due to the different underlying mechanisms of these three network models, achieving this is not trivial. The difficulty is encountered mainly with scale-free and small-world network models, and we discuss it further in Section~\ref{Network Generation Approach}. To obtain comparable results we have developed a set of mathematical formulas that allow generating networks following these three models with a particular number of vertices and edges. This, together with a network simulator that can be easily extended to include other network models and properties, is the primary contribution of our work. The developed simulator, together with generated networks have been open-sourced on GitHub as an R package\footnote{https://github.com/AkandaAshraf/netsim}. The entire simulation workflow has been automated, so the user only needs to define input parameters to get a comprehensive set of results in the shape of multiple plots and tables. The second and equally important contribution of this research is the comprehensive analysis of the results of over 30,000 simulations, run to generate and investigate networks following different network models.
We have measured three types of network properties: 1)~centralities (betweenness and closeness), 2)~average shortest path, and 3)~global clustering coefficient. For a given network model, we varied the number of vertices and edges, yet keeping them fixed across all three network types to make the results comparable.
In Section~\ref{Brief Overview of Network Generators Landscape} we present a brief overview of existing network generators. Next, the proposed \textbf{NetSim framework} together with implemented network models are introduced in Section~\ref{Proposed NetSim Framework}. In Section~\ref{Design of the Experiment}, we describe the design of the experiment that enables comparison of properties of implemented models, followed by results analysis in Section~\ref{Results and analysis}. Finally, conclusions are provided in Section~\ref{Conclusions}.

\section{Brief Overview of Network Generators Landscape}
\label{Brief Overview of Network Generators Landscape}

To the best of our knowledge, there does not exist a framework which generates comparable simulated networks in a comprehensive way for the random graph, scale-free, and small-world networks with a wide range of parameters but keeping the number of edges and vertices equal for all networks.

The \emph{igraph} package~\cite{csardi2006igraph} is a collection of network analysis tools available in R, Python and C. In our developed framework we have used \emph{igraph} library in R for generating individual models and calculating network properties. However, our NetSim framework allows to generate all three models, with constant number of edges and vertices for a range of different parameters and to save all the calculated properties as serialised objects which are then used to present the results in a transparent manner with generated plots. The whole process is automated and the user does not need to worry about keeping the number of edges and vertices the same, only vertices are taken as inputs.  

\emph{Brain Connectivity Toolbox}~\cite{rubinov2010complex} is a MATLAB toolbox for complex network analysis of structural and functional brain connectivity datasets. This toolbox has several models, including random and scale-free networks, and includes tools for network comparison which however focus mainly on different community structures, not network properties. 

\emph{MatlabBGL}\cite{gleich2007matlab} is another MATLAB package for working with graphs. It is based on the Boost Graph Library and enables generation of simple models like random graphs and cycle graph~\cite{siek2001boost}. 

\emph{Stanford Network Analysis Platform (SNAP)}~\cite{leskovec2016snap} is a library written in C++ for high-performance large-scale complex network analysis. It allows to generate regular and random graphs. 

\emph{BRITE}~\cite{medina2001brite} is another network topology generator. The main focus of this generator is to simulate network that mimics the characteristics of the actual Internet network. However, some important features are not considered in BRITE generator as the main goal is to reflect internet topology. One such not implemented feature is the small-world network model which the authors have pointed out in the documentation~\cite{medina2001brite}.   

There exist a plethora of other software packages that allow generating networks. While those presented above are just a few examples, it is important to emphasise that, to the best of our knowledge, none of them enables to generate, in an automated way, networks that are easily comparable. Existing tools enable to generate networks of different models, but each model is used in isolation. If the user wants to create networks that can be compared, i.e. having the same number of nodes and edges regardless of their structure, they have to work out all parameters by themselves which, as presented in the next section, is not a trivial task. 

\section{Proposed NetSim Framework}
\label{Proposed NetSim Framework}
In this section, we introduce the network models implemented in the NetSim framework as well as our approach to generate networks with a consistent number of nodes/edges regardless of their structure. We have mathematically derived and empirically demonstrated our framework to be able to simulate and support analysis with varying parameters of three different network models. Our proposed framework, which has been turned into an open-source R package, NetSim, is extremely user-friendly. We can simulate large-scale networks once, calculate their properties and then save those network properties as serialised objects which can then later be fetched from hard-drive to generate plots and perform comparative analysis. This reduces the complexity of the process as generating a large scale network is computationally expensive. Our R package is based on \emph{igraph}~\cite{csardi2006igraph} and \emph{ggplot2}~\cite{wickham2011ggplot2}, where the former is used for the core graph calculations and the latter for generating high resolution plots.  

\subsection{Network models}

In our study, we have simulated three widely used network models: (1)~Barab\'{a}si-Albert model for the scale-free network~\cite{barabasi1999emergence}, (2)~Watts-Strogatz small-world model for the small-world network~\cite{watts1998collective}, and (3)~Erd\H{o}s-R\'{e}nyi model for the random graph network~\cite{solomonoff1951connectivity,erdos1959random,erdos1960evolution,erdHos1961strength}. A brief introduction of these models is given below. 

\textbf{Random Graph:} In the random graph network model, one creates a network with some properties of interest (specific degree distribution) and otherwise random. Although random graph model was first studied by~\cite{solomonoff1951connectivity} this model is mainly associated with Paul Erd\H{o}s and Alfr\'{e}d R\'{e}nyi~\cite{erdos1959random,erdos1960evolution,erdHos1961strength}. The random graph model could also be thought of as choosing uniformly a network from the set of all possible networks with exactly $n$ vertices and $m$ edges. This model is often referred to as $G(n,m)$ model for networks where $n$ is the number of nodes and $m$ being the number of edges. A slightly different definition is $G(n,p)$ model where $p$ defines the probability of edges appearing between all possible pairs of vertices. 

\textbf{Scale-free Model:} The scale-free model shows power law node degree distribution $P(k) \sim k^{-\alpha}$ (where $k$ -- node degree and $\alpha$ -- parameter whose value is typically in the range $2 < \alpha  < 3$) which usually indicates some interesting underlying process~\cite{mitzenmacher2004brief,newman2005power}. This kind of distribution was first discussed by Price~\cite{price1976general}. Price, in turn, was inspired by Herbert Simon, who discusses power law in a variety of non-network economic data~\cite{simon1955class}. Simon showed mathematically the fact that `rich get richer' effect results in power law distribution. He called this mechanism `cumulative advantage', but it is more often known as `preferential attachment' as coined by Barab\'{a}si and Albert~\cite{barabasi1999emergence}. 

\textbf{Small-world Model:} Transitivity measured by the network clustering coefficient is one of the least understood properties in network analysis~\cite{newmannetworks}. Another important property we see in real networks is the small-world effect -- all nodes are connected with each other by relatively short paths. To model these two properties Watts and Strogatz introduced a small-world network model~\cite{watts1998collective}. Classical small-world model rewires edges in a simple circle model to random positions. In a slightly modified version of the model, the whole process starts from a circle and no edges are removed, but new edges are inserted between randomly chosen vertices~\cite{newman1999scaling}. There is an important parameter in this model $p$ -- the probability of rewiring/adding edges.

\subsection{Network Generation Approach}
\label{Network Generation Approach}
To compare properties of the three network models, two parameters have been chosen to be fixed. These are the number of edges $m$ and the number of vertices $n$. For the random graph model, it is simple as we use the $G(n,m)$ model and generate the ensemble directly.

In case of a scale-free network, because this is a growing network model, the final number of edges in the network depends on how many edges are attached in each step $S$ as well as on the total number of vertices $n$.  

The total number of edges in a scale-free network with $n$ nodes and $S$ edges added in each growth step is: 
\begin{equation}
   m = n \cdotp S- \frac{S\cdotp(S+1)}{2}. 
   \label{eq:scalefreeEdges}
\end{equation}
For a small-world network, the total number of edges for a network with $n$ vertices and $Nei$ number of neighbourhood within which the vertices of the lattice will be connected are:
 \begin{equation}
   m = n \cdotp Nei. 
    \label{eq:smallworldEdges}
\end{equation}
To keep the number of edges equal to $m$ for both the small-world network and the scale-free network (i.e. same as in Equation~\ref{eq:scalefreeEdges}) we randomly delete a fixed number of edges $x$ from the small-world network in Equation~\ref{eq:smallworldEdges}. As a result, the total number of edges for a small-world network is: 

\begin{equation}
 m = n \cdotp Nei - x. 
  \label{eq:smallworldEdgesX}
\end{equation}

If we now solve both Equations~\ref{eq:scalefreeEdges} and~\ref{eq:smallworldEdgesX} for $x$: 
\begin{equation}
\label{eq:solveForXFinal}
x = n\cdotp(Nei-S) +\frac{S\cdotp(S+1)}{2}
\end{equation}
From Equation~\ref{eq:solveForXFinal}, we can see that number of deleted edges does not depend on the number of vertices when $Nei$ and $S$ are equal as the first part of the equation on the right-hand side becomes $0$. Due to this, in our comparison, we have considered $S$ and $Nei$ parameters to be equal. When it is $0$, $\frac{S(S+1)}{2}$ might not have an impact on the type of the model for a large number of vertices as long as $S$ is comparatively small. In our comparisons, we have considered $S$ up to $16$. The value is chosen as power of $2$, where the power is from $1$ to $4$ i.e. $S \in \{2,4,8,16\}$.

We have generated networks for scale-free and random graphs with $n$ vertices and $m$ edges. The number of edges is calculated from Equation~\ref{eq:scalefreeEdges}, which is the number of edges we get in scale-free networks for a different number of vertices $n$. For small-world networks, first, a lattice with $n$ vertices and $Nei$ number of neighbourhood within which the vertices of the lattice will be connected is generated. From the generated lattice, $x$ edges (Equation~\ref{eq:solveForXFinal}) are randomly removed. After that, the edges are rewired with probability $p$ to obtain small-world network.

Following the process described above, we obtained networks of the three discussed models that are comparable as they have the same number of nodes and edges which, in turn, enables comparative analysis of different networks' properties including clustering coefficient, different types of centralities and average shortest paths.

\section{Design of the Experiment}
\label{Design of the Experiment}
We have varied a number of parameters to generate different networks within three models. For all three network types we have $n\in\{100,200,300,400,500,600,700,800,900,1000,2000,3000,4000,5000,6000,7000,8000,\\9000,10000\}$ vertices. For scale-free networks $\alpha~\in \{ 1.5,  1.75, 2, 2.25, 2.5, 2.75, 3, 3.25, 3.5 \}$ and for small-world networks $p \in \{ 0.3, 0.4, 0.5, 0.6, 0.7 \}$ are used. 

The number of edges for scale-free and small-world networks is calculated using Equations~\ref{eq:scalefreeEdges} and~\ref{eq:smallworldEdgesX} respectively. For the random graph, we have used $G(n,m)$ model where the number of edges is directly defined. Note, that number of edges will be the same for all networks for a given set of $S$ and $Nei$ values. Table~\ref{tab:1} shows the number of edges which correspond to each $S$ and $Nei$ parameters for each vertex count defined from set $n$. 

\begin{table}

\caption{\label{tab:1} Number of edges for different values of $S$ and $Nei$}
\begin{tabular}{lll}
\hline
$S$ & $Nei$ & Number of edges in a network                                                                                                                                            \\ \hline
2    &2         & \begin{tabular}[c]{@{}l@{}}197, 397, 597, 797, 997, 1197, 1397, 1597, 1797, 1997, 3997, 5997, 7997, 9997, \\11997, 13997, 15997, 17997, 19997\end{tabular}                  \\ 
4      &4       & \begin{tabular}[c]{@{}l@{}}390, 790, 1190, 1590, 1990, 2390, 2790, 3190, 3590, 3990, 7990, 11990, 15990, \\19990, 23990, 27990, 31990, 35990, 39990\end{tabular}            \\ 
8        &8     & \begin{tabular}[c]{@{}l@{}}764, 1564, 2364, 3164, 3964, 4764, 5564, 6364, 7164, 7964, 15964, 23964, 31964, \\39964, 47964, 55964, 63964, 71964, 79964\end{tabular}          \\ 
16         &16   & \begin{tabular}[c]{@{}l@{}}1464, 3064, 4664, 6264, 7864, 9464, 11064, 12664, 14264, 15864, 31864, 47864, \\63864, 79864, 95864, 111864, 127864, 143864, 159864\end{tabular} \\ \hline
\end{tabular}
\end{table}

In our simulation, we have shown empirically that our method keeps the number of edges and vertices the same for all types of networks and is consistent with Equations~\ref{eq:scalefreeEdges} and~\ref{eq:smallworldEdgesX}.



For each combination of $\alpha$ and $p$, both $S$ and $Nei$ are increased as a power of 2. $S,Nei \in \{2,4,8,16\}$. The reason behind keeping $S$ and $Nei$ the same is described in Section~\ref{Network Generation Approach}. Each of the networks was generated 30 times, and the mean for each analysed property was calculated over those 30 samples. Table~\ref{tab:table-totalNumberOfGenNet} summarises the total number of generated networks. Without sampling, we have in total 1,083 different types of networks with different parameters. 
\begin{table}
    \caption{\label{tab:table-totalNumberOfGenNet} Number of simulated networks.}
    \begin{tabular}{llllll}
    \hline
    Network Model & $\mid n \mid$ & $ \mid \alpha \mid$ & $ \mid p \mid$    & $\mid S \mid$ or $\mid Nei \mid$        & Sampling               \\  \hline
    Scale-free   & 19           & 19 $\times$  9 = 171 & & 171 $\times$ 4 =  684 & 684 $\times$ 30 = 20520 \\ 
    Small-world  & 19           & & 19 $\times$ 5 = 95 &95 $\times$ 4 =  380  & 380 $\times$ 30 =11400   \\ 
    Random Graph & 19           & 19                   & &19                    & 19 $\times$ 30 = 570      \\ 
     \hline 
    ~            & ~            &  ~                     &Total   & 1083                     & 32490                      \\  \hline
    \end{tabular}
\end{table}

\section{Results and analysis}
\label{Results and analysis}

In this section, we present results that were obtained using the NetSim framework for five different properties and three different network models with varying size and density.

\subsection{Number of Edges and Vertices}

\begin{figure}[!ht]
\centering
\includegraphics[width=0.3\textwidth]{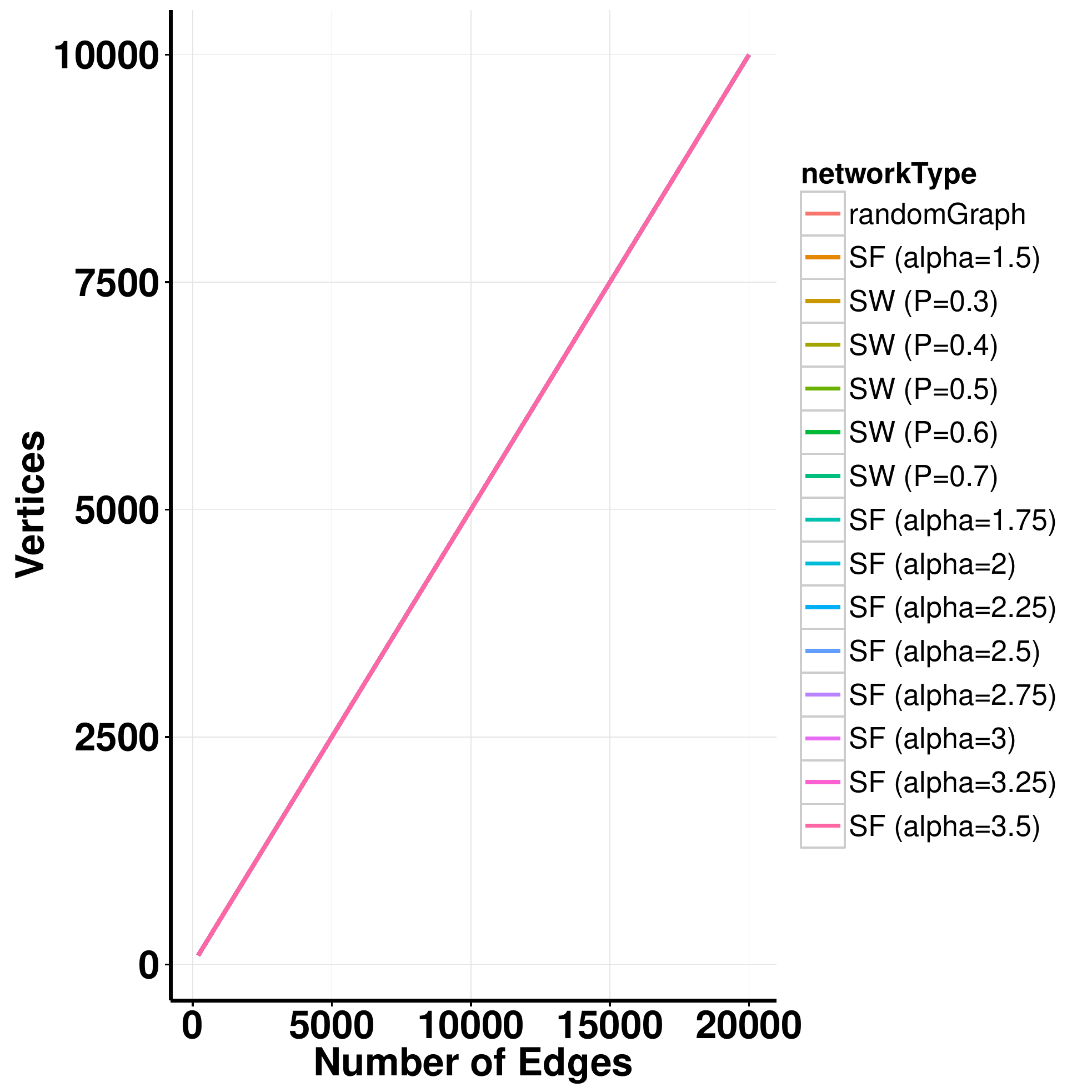}

\caption{Edge and Vertices plot for random graph networks, small-world networks, and scale-free networks; $S=2$ $Nei=2$.}

\label{fig:edgesAndVerticesLine}
\end{figure}

Figure~\ref{fig:edgesAndVerticesLine} shows the number of edges in relation to the number of vertices for different networks. Plots for all networks overlap and form a straight line. This is because in each of the networks we have the same number of vertices and edges which shows empirically that our proposed method described in Section~\ref{Network Generation Approach} is viable. In Figure~\ref{fig:edgesAndVerticesLine} the number of edges in relation to the number of vertices is presented for $Nei=2$ and $S=2$ and the same trend holds for all other experimental settings from Table~\ref{tab:1}.

\subsection{Closeness Centrality}

\begin{figure}[!ht]
\centering
\subfigure[$S$ \& $Nei$ =$2$]{\label{fig:a}\includegraphics[width=60mm,height=54mm]{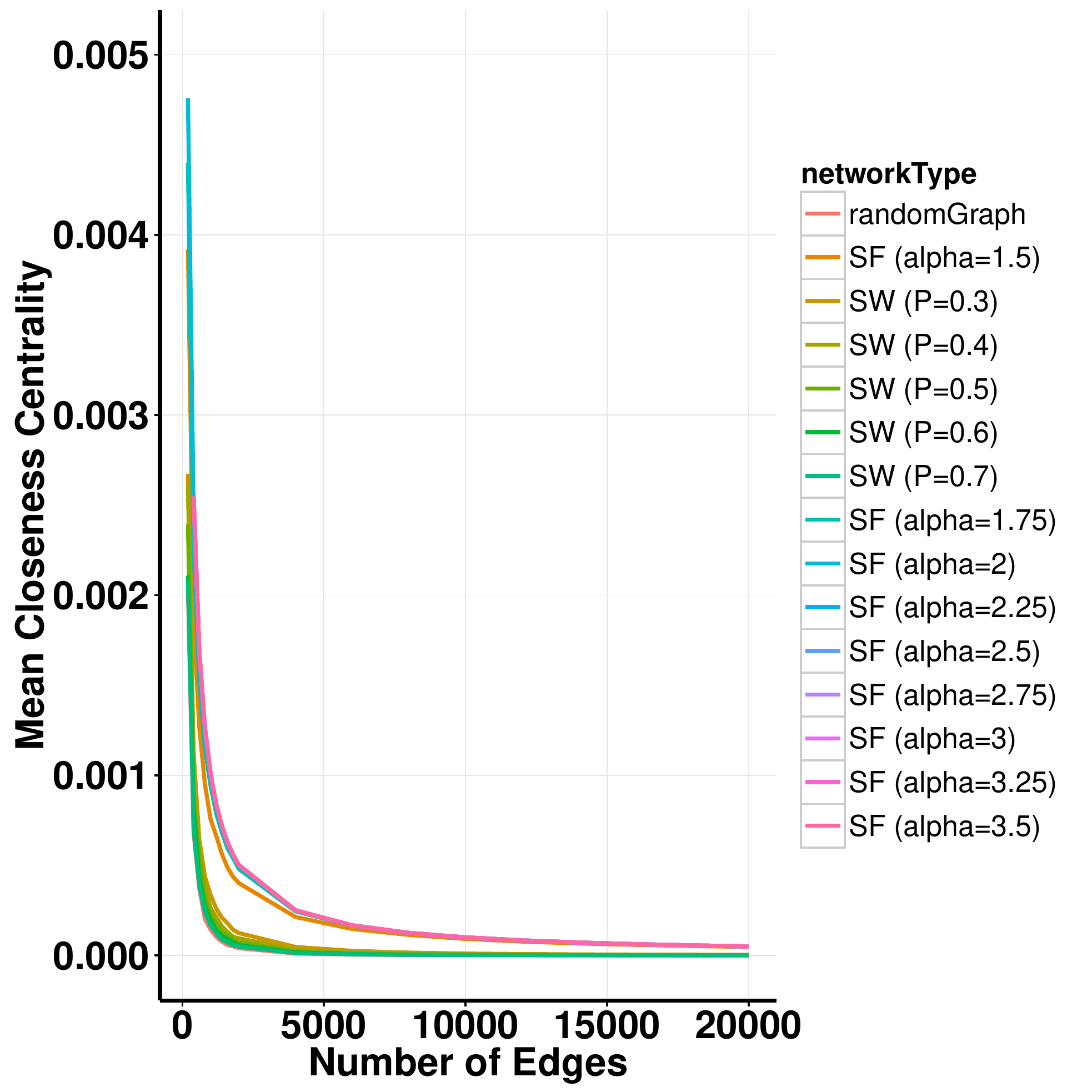}}
\subfigure[$S$ \& $Nei$ =$4$]{\label{fig:b}\includegraphics[width=60mm,height=54mm]{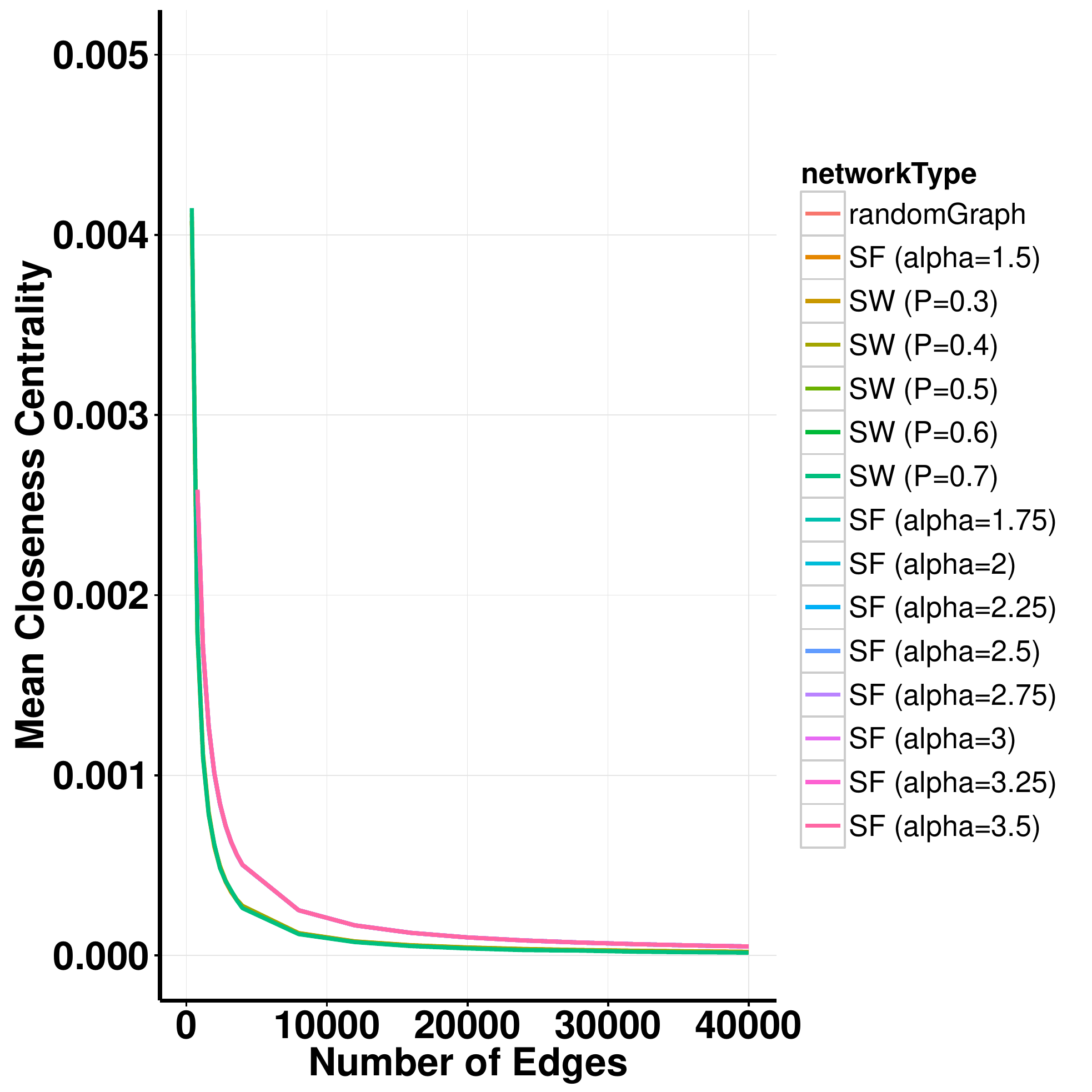}}
\subfigure[$S$ \& $Nei$ =$8$]{\label{fig:b}\includegraphics[width=60mm,height=54mm]{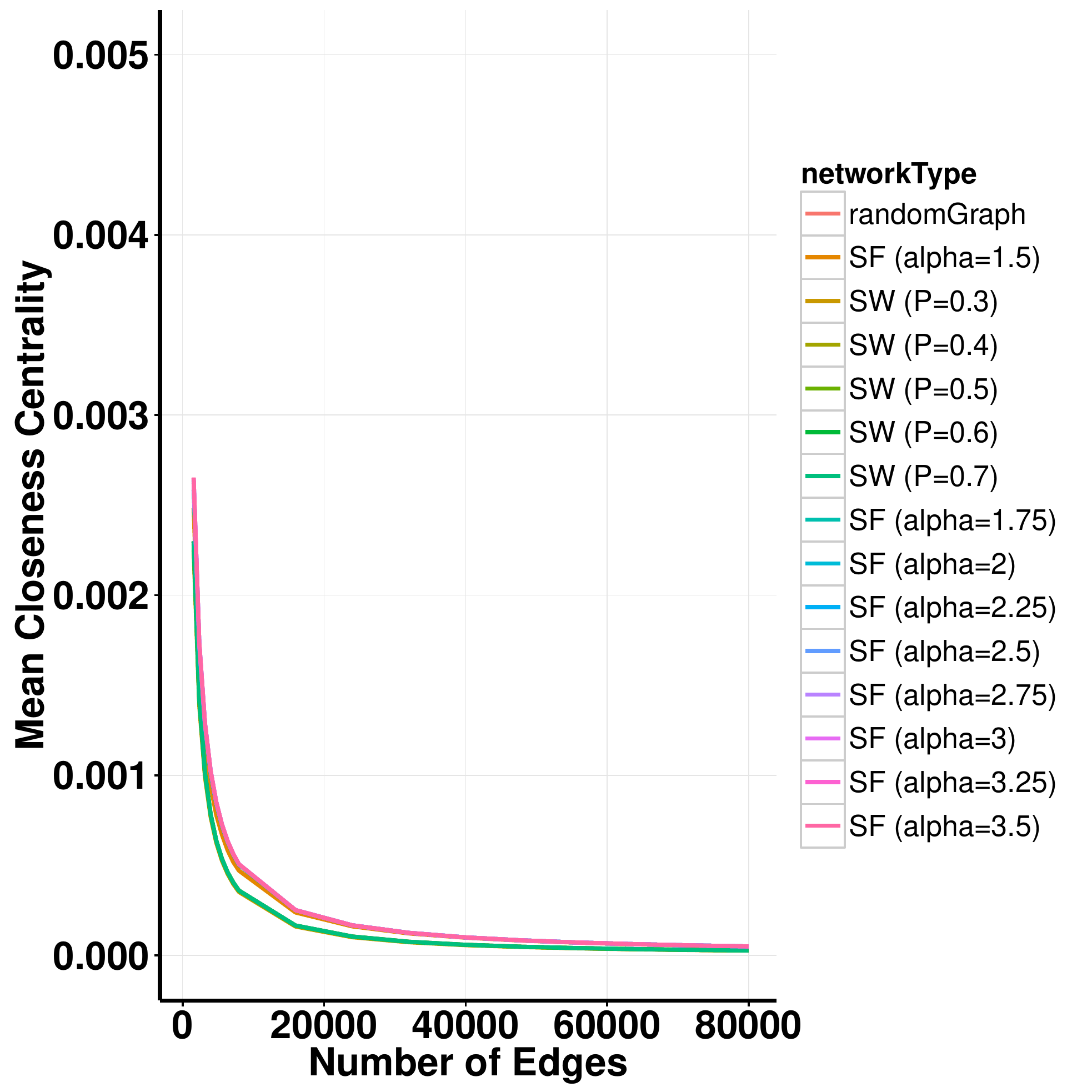}}
\subfigure[$S$ \& $Nei$ =$16$]{\label{fig:b}\includegraphics[width=60mm,height=54mm]{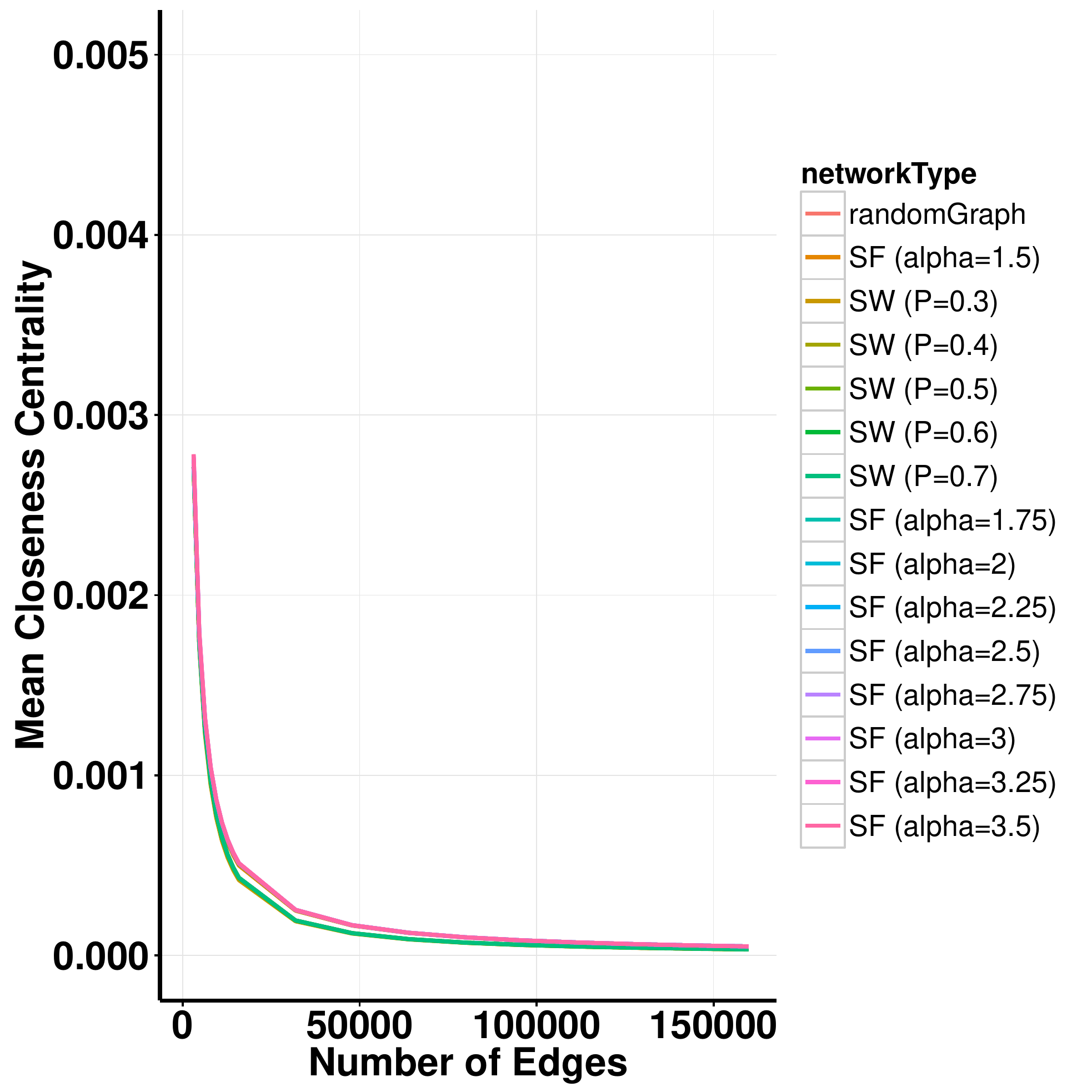}}
\caption{Closenesss Centrality in relation to edges for random graph, small-world, and scale-free networks with different values of $S$ and $Nei$.}
\label{fig:edgesAndClosenessLine}
\end{figure}

Figure~\ref{fig:edgesAndClosenessLine} shows mean closeness centrality in relation to the number of edges. Closeness centrality is expected to have higher values in social networks due to the small-world effect, which implies that every vertex should be connected with other vertices via a short path even if the network is large. 

As network size is increased closeness centrality decreases for all types of networks. This is because all generated networks become sparser as they grow in terms of the number of vertices (i.e. moving from left to right of a single plot). Although we have increased the number of vertices and edges together, the number of edges is not increased sufficiently with respect to vertices to maintain a dense network. However, increasing the value of $S$ and $Nei$ implies more edges, thus a denser network. For the value of $16$ of both $Nei$ and $S$, the simulation results in networks with the biggest number of edges among all generated networks. For $Nei=16$ and $S=16$  mean closeness centrality is similar for all generated networks. This implies that as the network density grows, different types of networks might exhibit similar behaviour in terms of closeness centrality.

When comparing different network types, one can see that random graph has the lowest value for mean closeness centrality. This is something expected as a purely randomly grown network does not have the small-world effect. Scale-free networks with higher values of $\alpha$ seem to have a higher value of mean closeness centrality. This might, in turn, imply that the `rich get richer' effect has a positive impact on the mean closeness centrality of the entire network.  

\subsection{Betweeness Centrality}

\begin{figure}
\centering
\subfigure[$S$ \& $Nei$ =$2$]{\label{fig:a}\includegraphics[width=60mm,height=54mm]{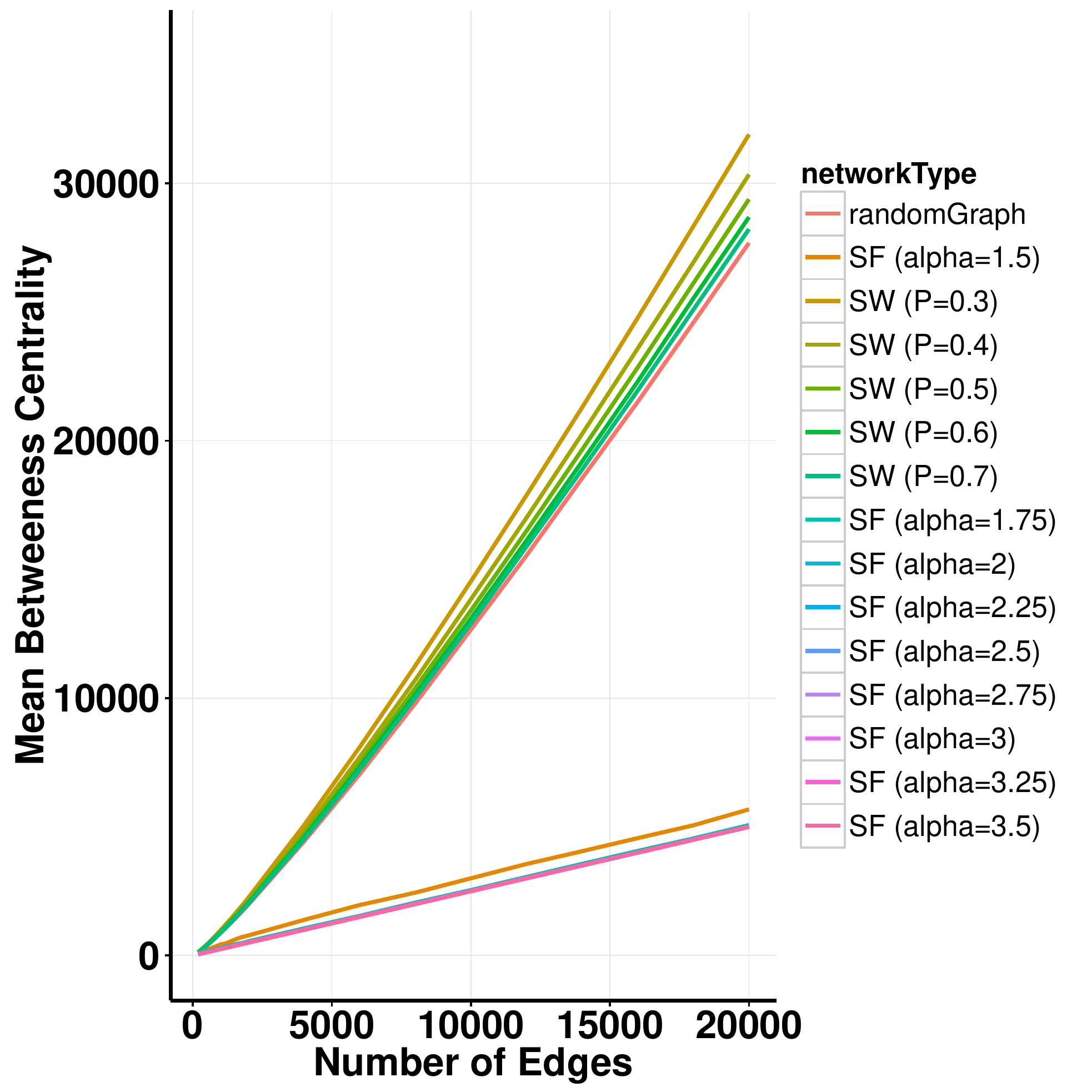}}
\subfigure[$S$ \& $Nei$ =$4$]{\label{fig:b}\includegraphics[width=60mm,height=54mm]{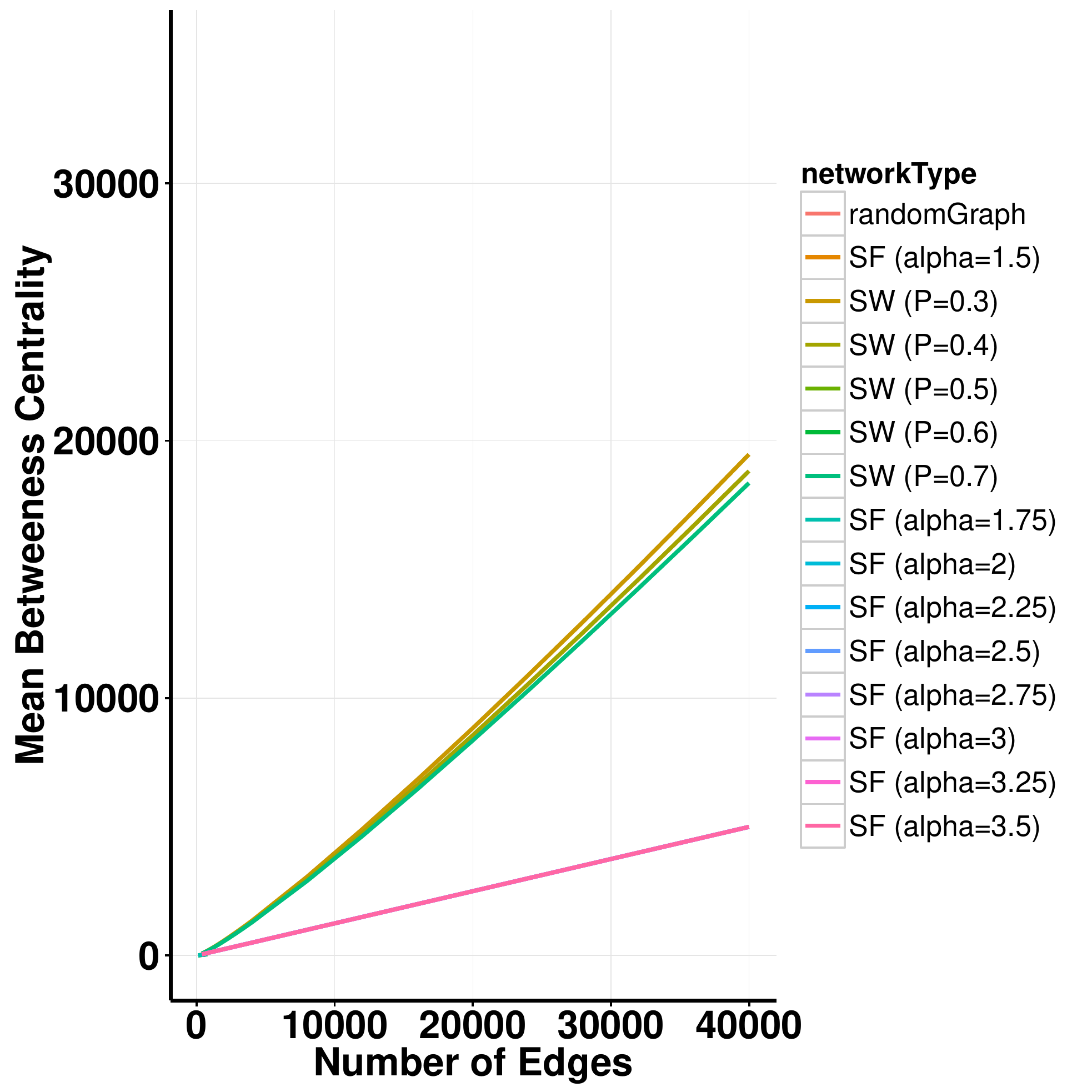}}
\subfigure[$S$ \& $Nei$ =$8$]{\label{fig:b}\includegraphics[width=60mm,height=54mm]{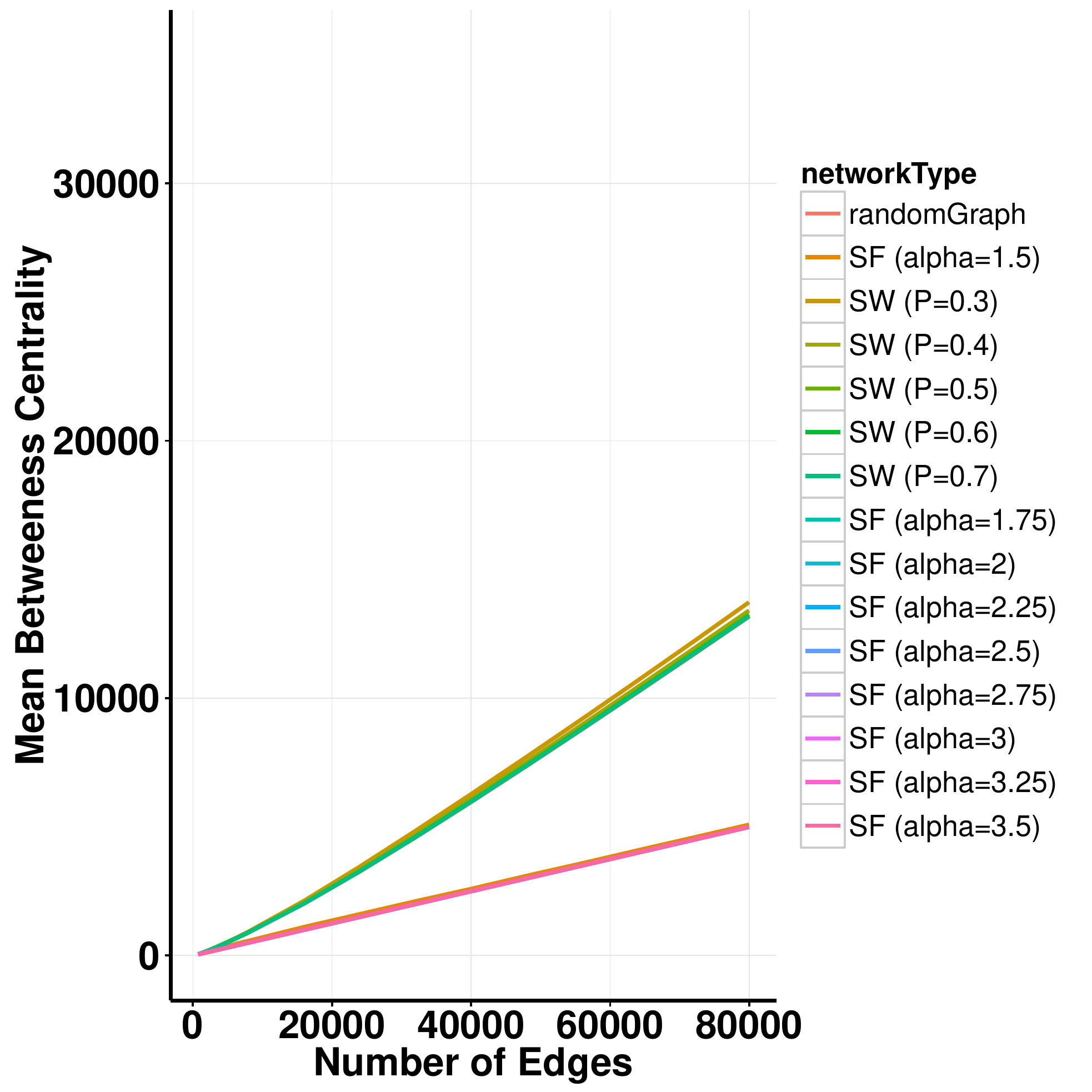}}
\subfigure[$S$ \& $Nei$ =$16$]{\label{fig:b}\includegraphics[width=60mm,height=54mm]{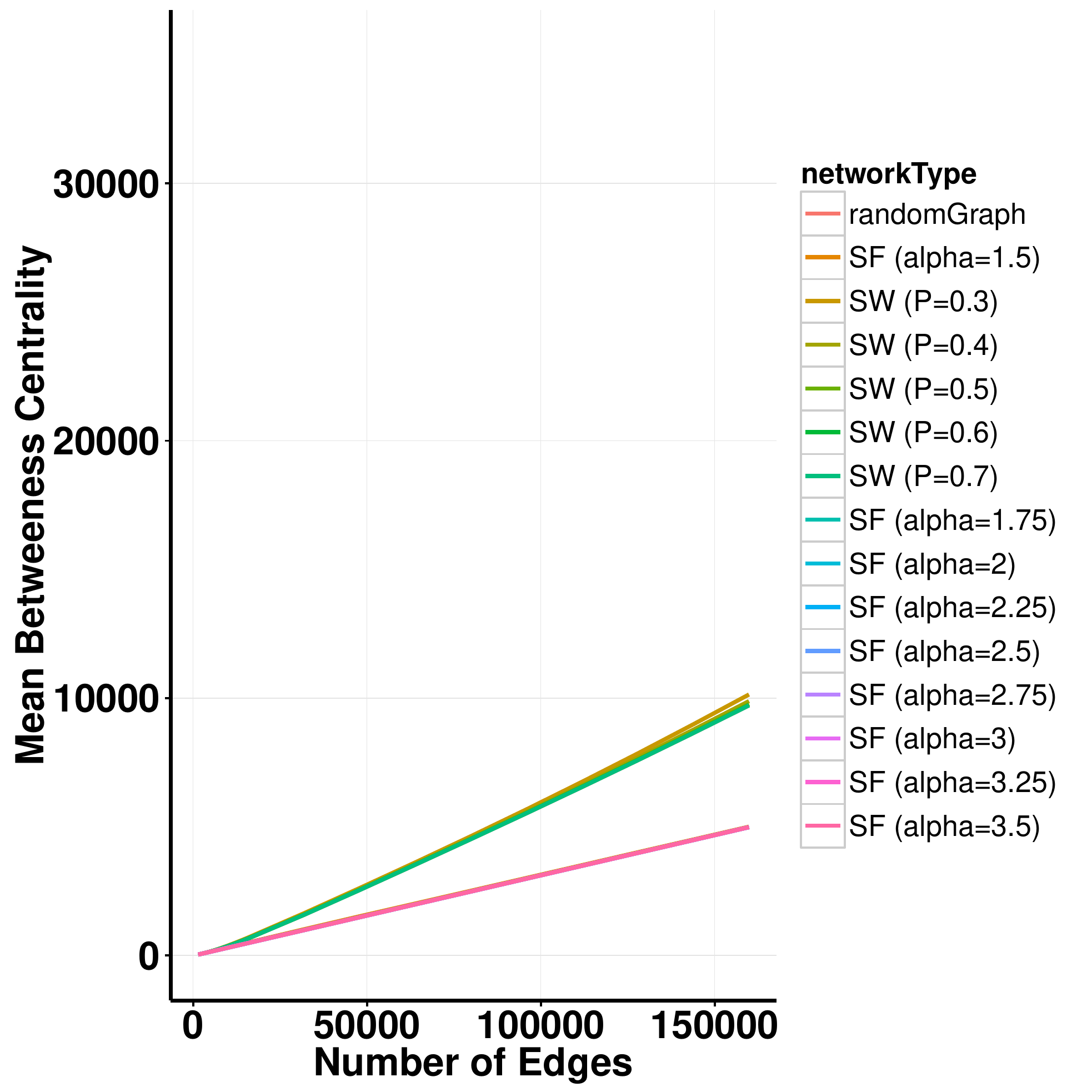}}

\caption{Betweenness Centrality in relation to edges for random graph, small-world, and scale-free networks with different values of $S$ and $Nei$.}
\label{fig:edgesAndBetweenessLine}
\end{figure}

Betweenness centrality is a measure of how much influence a vertex has on the information flow within a network. We see a linear increase in the mean betweenness centrality values for all the networks with increasing size of the network (Figure~\ref{fig:edgesAndBetweenessLine}). This is because as a network becomes bigger regarding the number of vertices, each vertex lies between more vertices in general. 

One important observation here is that there are two clusters of networks in each of the plots. There is a clear difference between scale-free networks and small-world/random networks. As we make networks denser by increasing values for $S$ and $Nei$, networks in those two clusters remain separate but tend to behave more like each other within the same cluster. We also see that as the network gets denser, betweenness centrality decreases.

Scale-free networks tend to have lower values of betweenness centrality, and the value decreases with growing values of $\alpha$. This can imply that the `rich get richer' effect is responsible for the decrease in mean betweenness centrality values. Although for small-world and random networks this centrality measure is far higher than scale-free networks, randomness tends to play a negative role here. We see that with small-world networks, higher values of $p$ tend to have lower mean betweenness centrality although they lie in the upper cluster. 

\subsection{Average Shortest Path}

\begin{figure}[!ht]
\centering
\subfigure[$S$ \& $Nei$ =$2$]{\label{fig:a}\includegraphics[width=60mm,height=54mm]{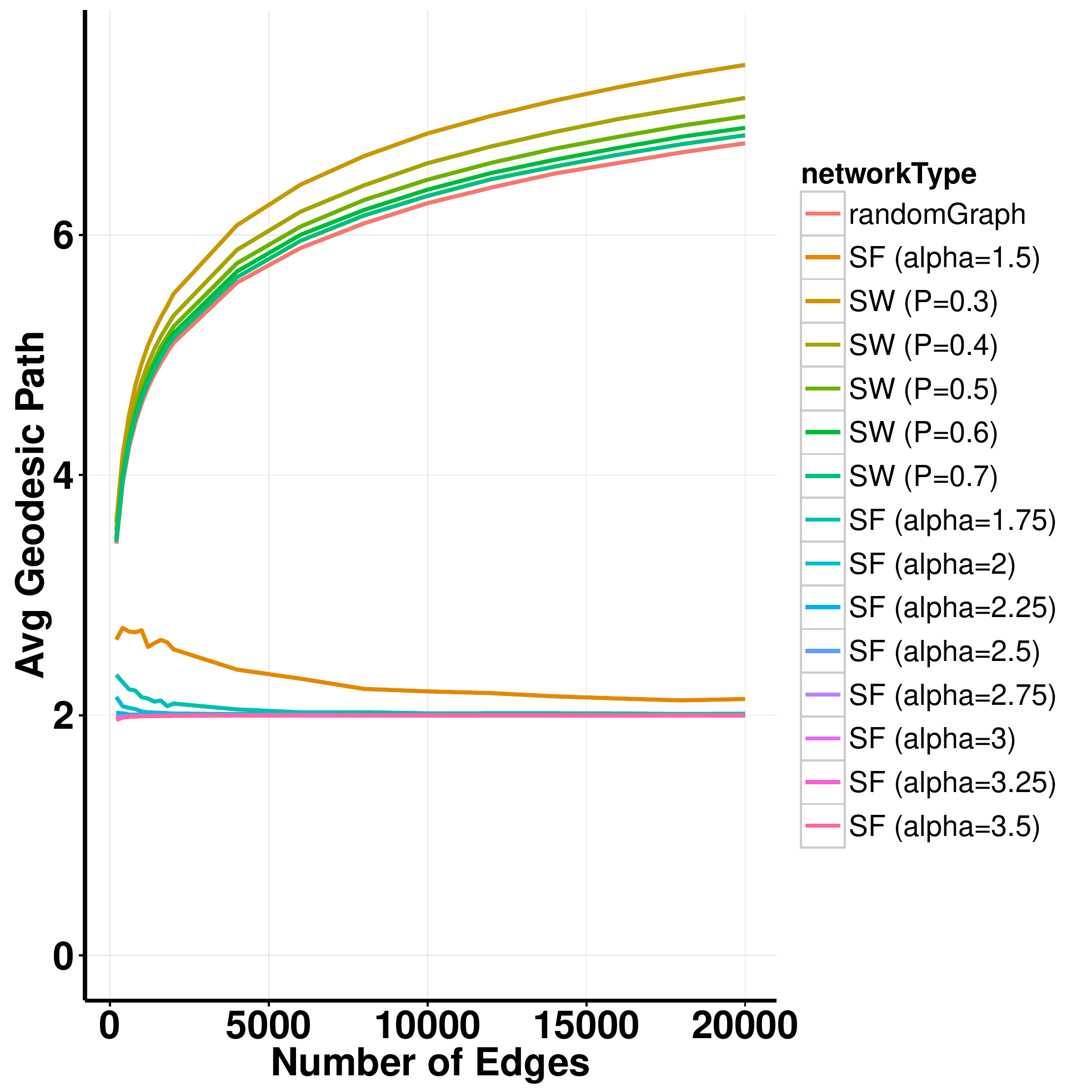}}
\subfigure[$S$ \& $Nei$ =$4$]{\label{fig:b}\includegraphics[width=60mm,height=54mm]{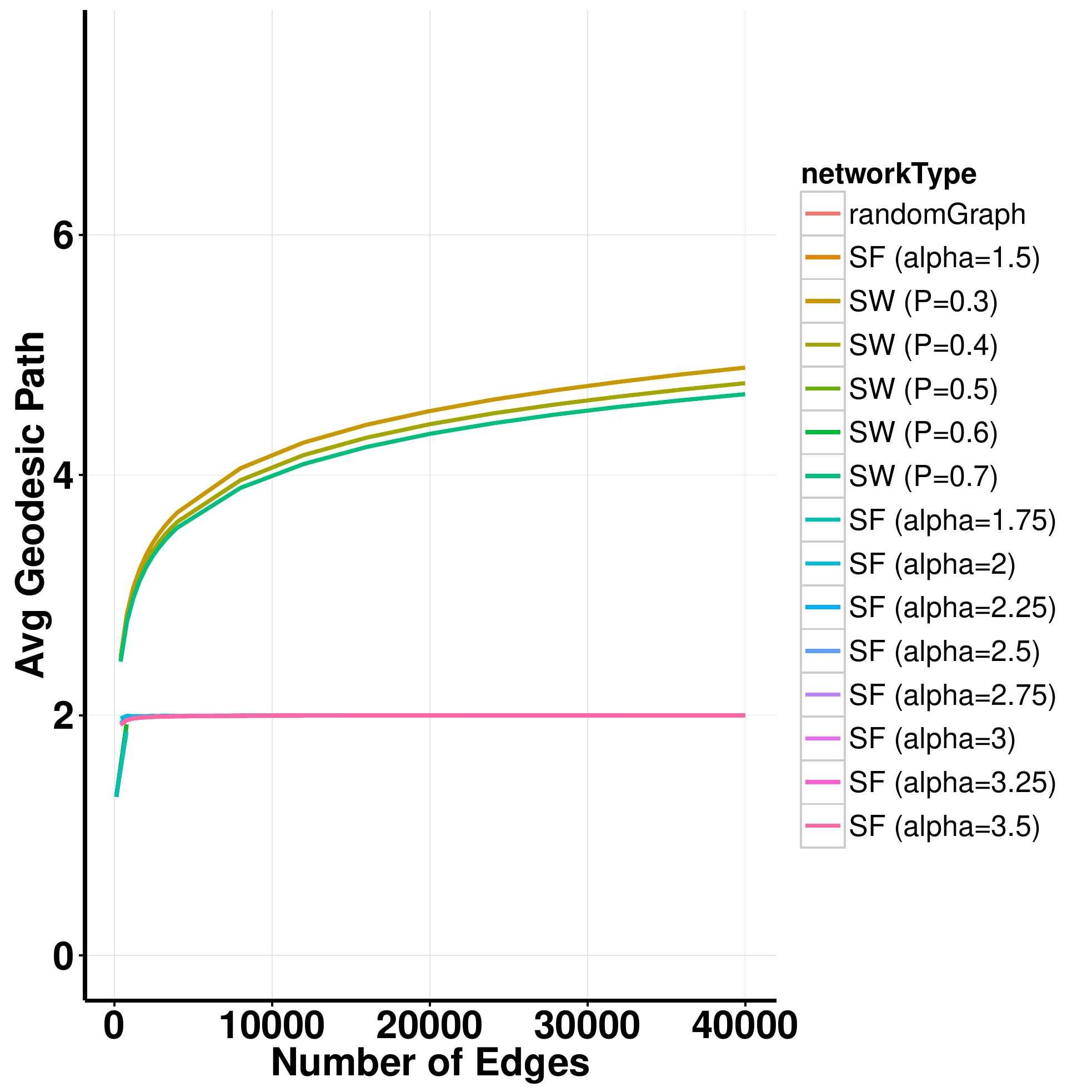}}
\subfigure[$S$ \& $Nei$ =$8$]{\label{fig:b}\includegraphics[width=60mm,height=54mm]{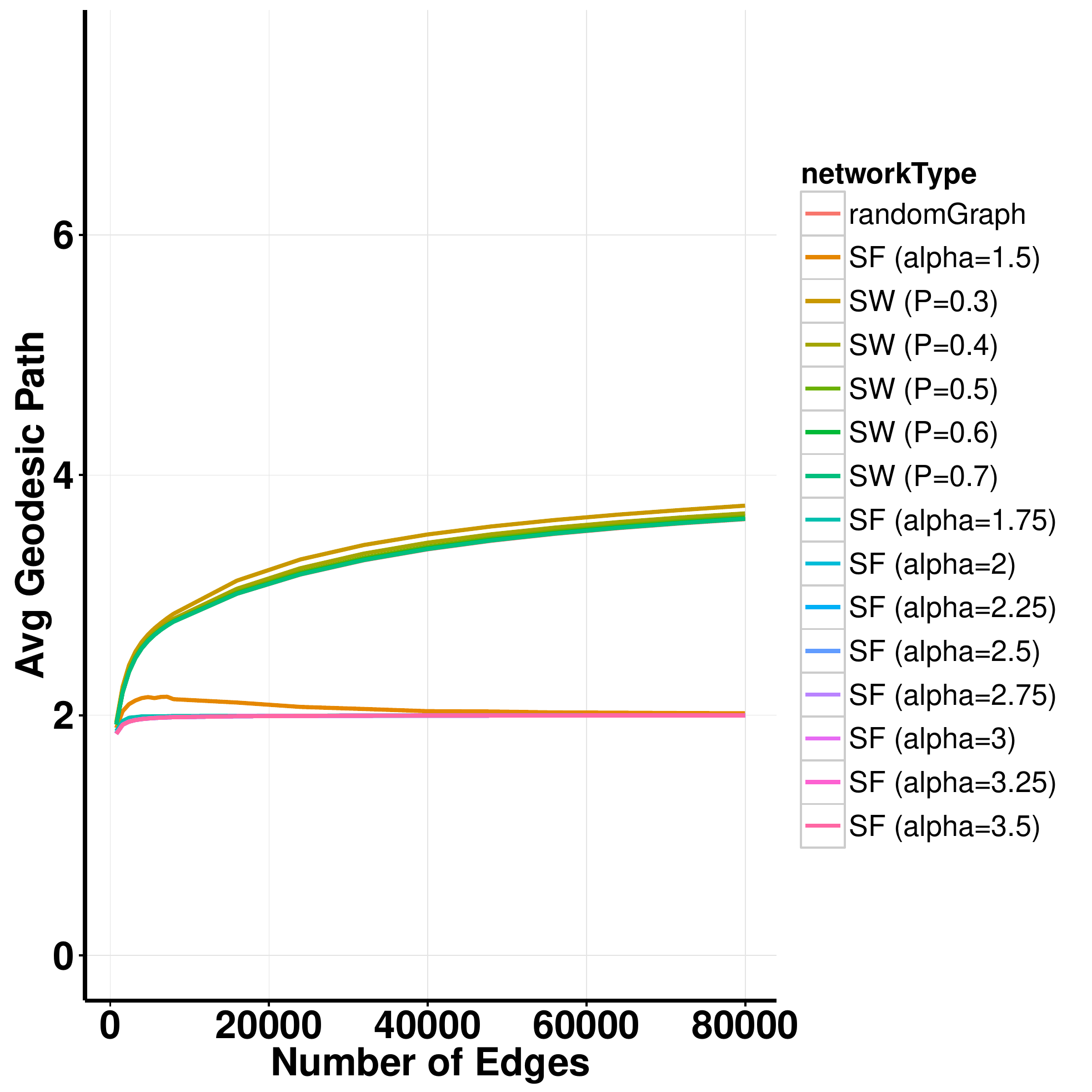}}
\subfigure[$S$ \& $Nei$ =$16$]{\label{fig:b}\includegraphics[width=60mm,height=54mm]{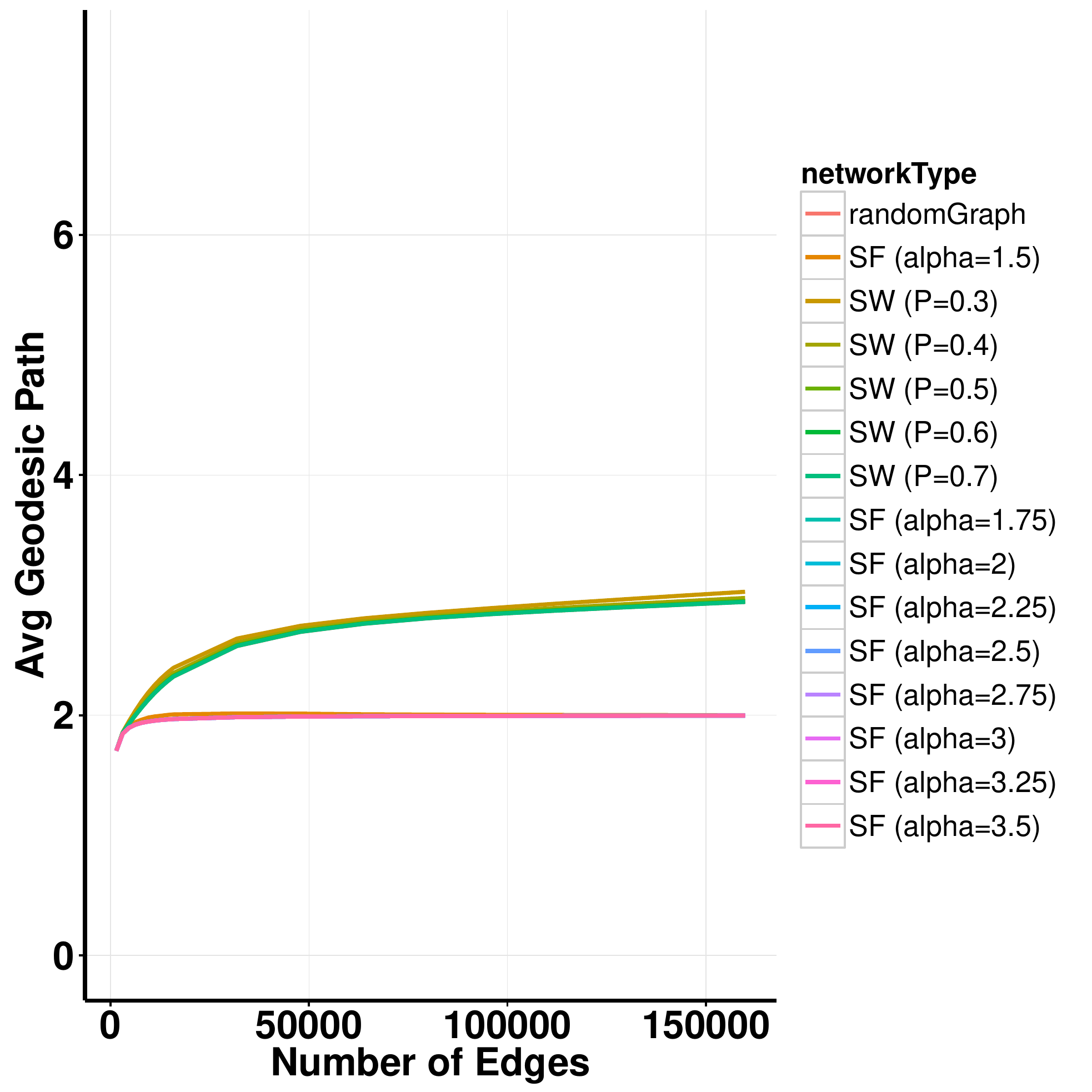}}

\caption{Average Shortest Path in relation to edges for random graph, small-world, and scale-free networks with different values of $S$ and $Nei$.}
\label{fig:edgesAndBetweenessLine}
\end{figure}

The average shortest path is calculated as the mean shortest path for all pairs of vertices. In our analysis, the average of shortest paths between all pairs of vertices is plotted against the number of edges in Figure~\ref{fig:edgesAndBetweenessLine}.

It is interesting to see that when $S$ and $Nei$ are increased, resulting in a denser network, the shortest path is decreased for all three types of networks with different parameters. This implies that the small-world phenomena are more apparent in a denser network and this is true for every small-world, random graph, and scale-free network. 
 

Small-world networks with lower values of $p$ tend to have longer average shortest paths, whereas networks with higher values of $p$ have shorter average shortest path. This is intuitive as having lower values of $p$ means fewer connections are rewired in our networks from the initial lattice which has a high clustering coefficient but lower intensity of the small-world phenomenon i.e. longer shortest path. When the value of $p$ is increased, it gets closer to a random network, which has a shorter average geodesic path. We can also see this phenomenon for the simulated random networks, which have a very short average geodesic path.

Interestingly, size of the network does not seem to influence the value of average geodesic/shortest path of scale-free networks, and they stay almost constant. Whereas, for the cases of small-world and random graph networks, as the number of edges and vertices increase, the average shortest path also increases. In fact, for some of the scale-free networks, increasing the size of the network tends to result in smaller values of the average geodesic paths. This is expected, as new edges tend to be attached to vertices where we already have edges due to the `rich get richer' phenomena. This results in a stable value of average geodesic path with regards to the size of the network. 
 
One important note for all the plots in Figure~\ref{fig:edgesAndBetweenessLine} is that as we go right on the $x$-axis, we have more edges and vertices as detailed in Table~\ref{tab:1}. However, with a higher number of vertices, the network tends to become sparser as we are not increasing the number of edges at a rate which will keep the density constant.  On a single plot, when we move from left to right, we have sparser and larger networks while in four plots with different parameters of $S$ and $Nei$, increasing the number of $S$ and $Nei$ results in denser networks. 

\subsection{Global Clustering Coefficient}

The global clustering coefficient (a.k.a. transitivity) is measured as the ratio of the triangles (for three vertices it includes three closed triplets) to the connected triples in networks.

\begin{figure}[!h]
\centering
\subfigure[$S$ \& $Nei$ =$2$]{\label{fig:a}\includegraphics[width=60mm,height=54mm]{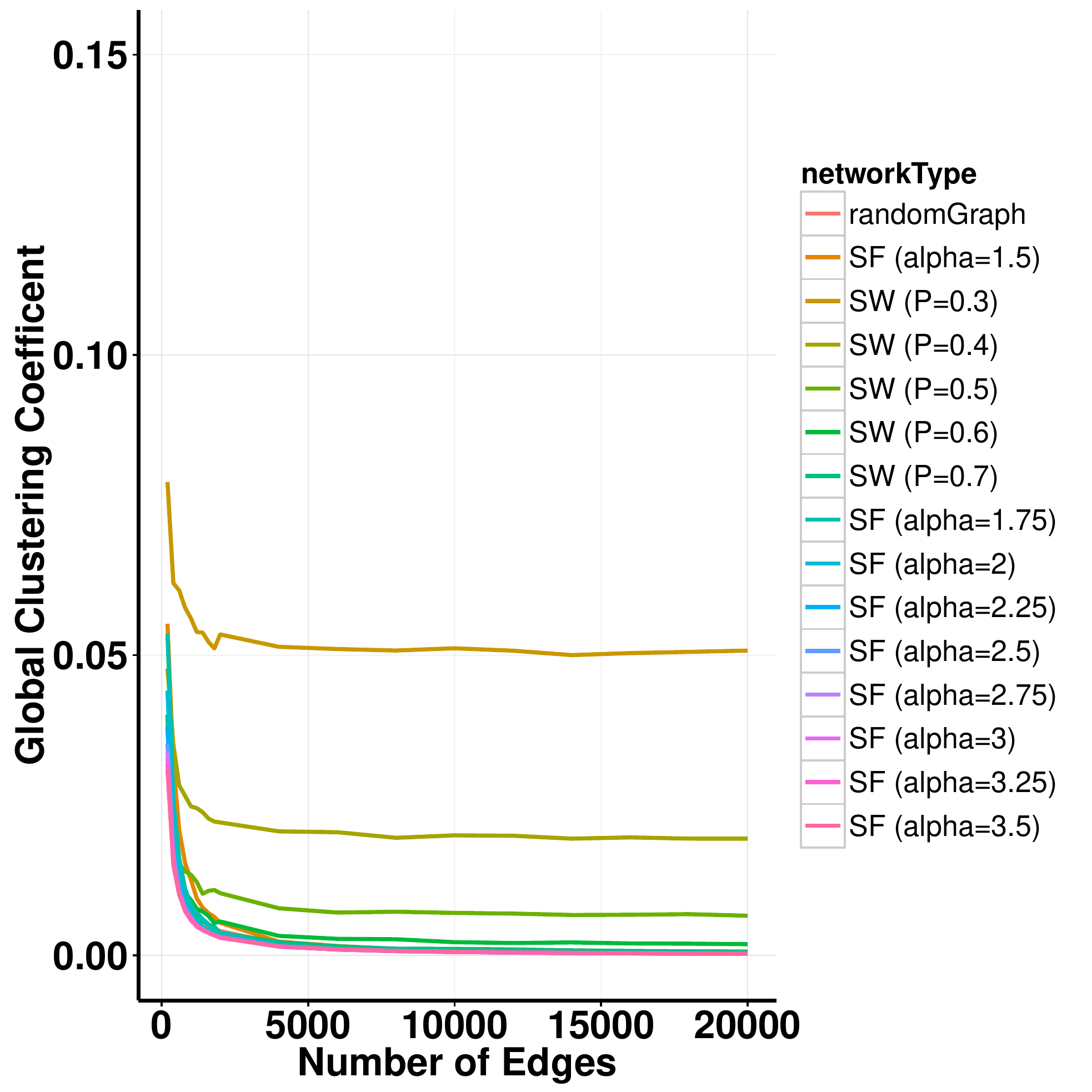}}
\subfigure[$S$ \& $Nei$ =$4$]{\label{fig:b}\includegraphics[width=60mm,height=54mm]{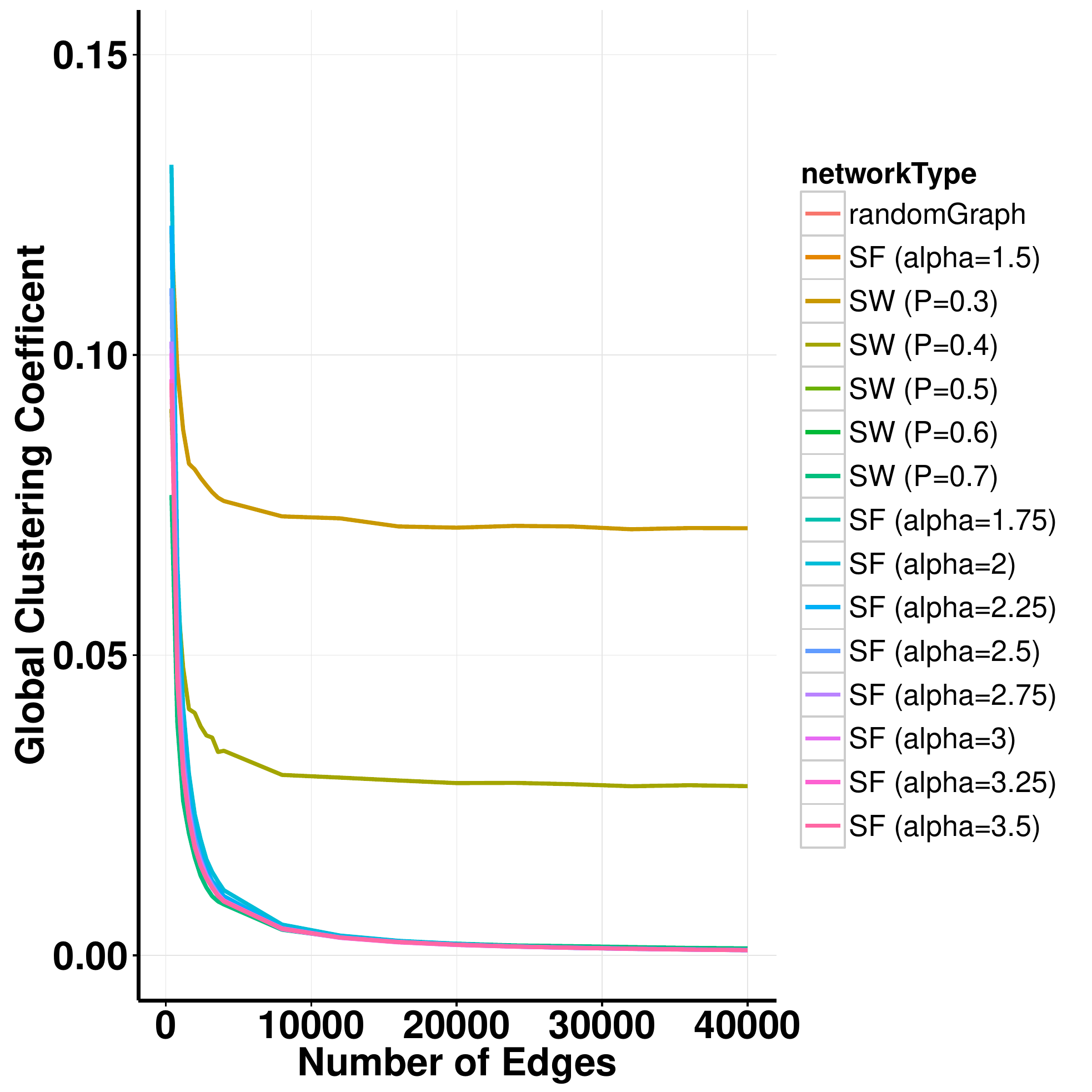}}
\subfigure[$S$ \& $Nei$ =$8$]{\label{fig:b}\includegraphics[width=60mm,height=54mm]{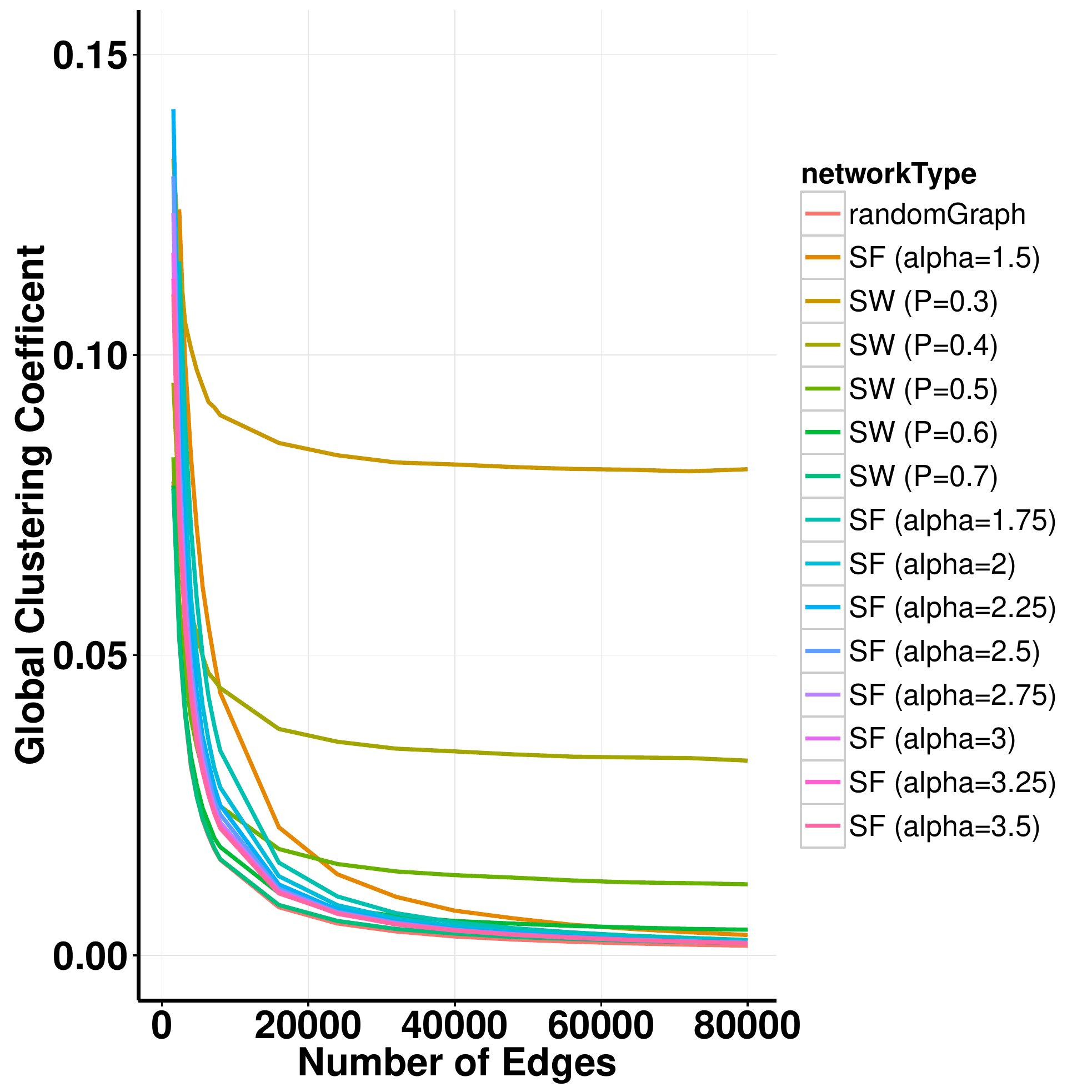}}
\subfigure[$S$ \& $Nei$ =$16$]{\label{fig:b}\includegraphics[width=60mm,height=54mm]{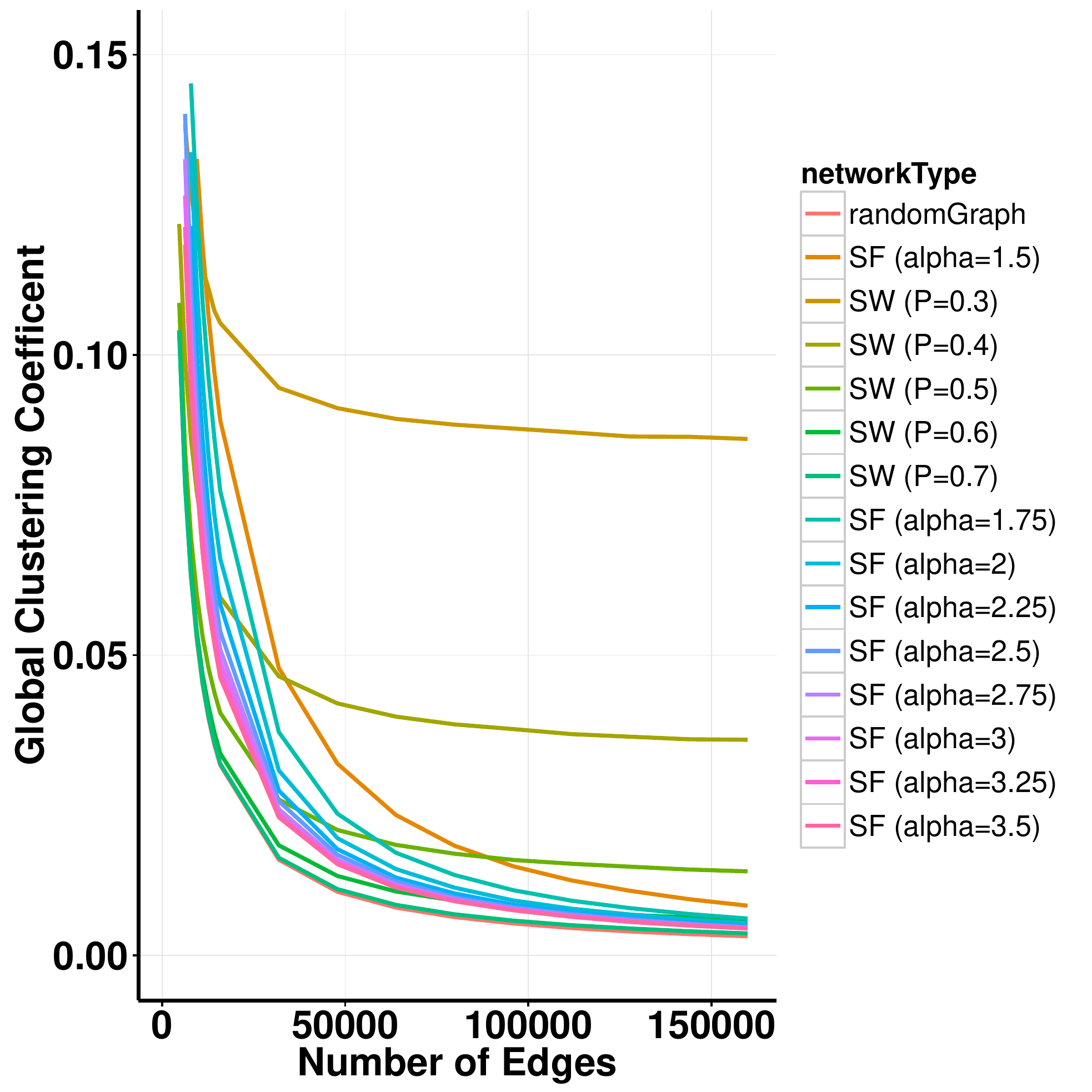}}

\caption{Global Clustering Coefficient in relation to edges for random graph, small-world, and scale-free networks with different $S$ and $Nei$.}
\label{fig:edgesAndCCLine}

\end{figure}

We observe larger values for transitivity for small-world networks with lower values of $p$, which is expected (Figure~\ref{fig:edgesAndCCLine}). The lower the values of $p$, the smaller the number of connections that are rewired, and the network is closer to the initial lattice which has a higher number of connected triples. Also, scale-free network tends to have lower values of transitivity and it decreases with increasing $\alpha$ (Figure~\ref{fig:edgesAndCCLine}). This is also expected as a scale-free network is biased towards already existing links, and they are not evenly spread, resulting in a lower number of connected triples. 
Another important observation is that transitivity seems to be constant with respect to increasing number of edges and vertices but not density. With denser networks (increasing $S$ and $Nei$) transitivity tends to decrease until it reaches a stable point. This is more visible with small-world networks that have a smaller number of edges rewired. Other networks seem not to be significantly influenced by the changes in density.

\section{Conclusions}
\label{Conclusions}
Our NetSim software allowed us to conduct a simulation study that resulted in comparative analysis of three main network models: random, small-world and scale-free. For all generated networks, we analysed closeness and betweenness centrality as well as average shortest path and global clustering coefficient.

Our experiments allowed to compare selected properties of different network models. The conducted study revealed some interesting insights into how different network structures influence the properties in question. 

Results for clustering coefficient and average shortest path confirmed the analytical results that are known for the three considered models~\cite{newmannetworks}. Additionally, it is apparent that for scale-free networks the average shortest path does not depend on network density, whereas for small-world and random networks the average shortest path decreases as the density increases. The clustering coefficient achieves the highest value, regardless of the density, for the small-world network with the smallest probability of rewiring $p=0.3$. This confirms the theoretical consideration as this network in its structure, is closest to regular lattice which has very high clustering.

For closeness centrality, especially for denser networks, there is not much difference across the network models. For sparse networks, closeness centrality tends to be smaller for small-world and random networks than scale-free networks. For scale-free networks, density does not influence closeness centrality, and it becomes stable. Whereas, for small-world and random networks the closeness centrality increases as the network gets denser. 
For betweenness centrality, there is a clear difference between scale-free networks and other models. Betweenness for scale-free networks is consistently smaller compared with other models. We attribute that to the `rich get richer' phenomenon that causes a concentration of edges in hub nodes. 
This study offers a generic network simulator and a set of analyses which have revealed some interesting, and previously unknown characteristics of the networks. Our results indicate that only by looking at a wide spectrum of generated networks we can identify extraordinary phenomena (e.g. mean betweenness centrality differentiates a scale-free network from a random graph or small-world network) that can otherwise be overlooked. Such information is useful not only to identify the type of real-world networks but also to calculate properties that can not be derived analytically.

\end{document}